\documentclass[12pt, letterpaper]{JHEP3}
\usepackage{latexsym,amssymb,amsmath,epsfig,epic,eepic}
\usepackage{epstopdf}
\newcommand{\beq}{\begin{equation}} \newcommand{\eeq}{\end{equation}}
\newcommand{\beqs}{\begin{eqnarray}}
\newcommand{\eeqs}{\end{eqnarray}} \newcommand{\tr}{\mathrm{tr\,}}
\newcommand{\Ncal}{\mathcal{N}} \newcommand{\Acal}{\mathcal{A}}
\newcommand{\Fcal}{\mathcal{F}} 
\newcommand{\Ccal}{\mathcal{C}}
\newcommand{\diag}{\mathrm{diag\,}}
\title{Stringy Instantons at Orbifold Singularities}

\author{\parbox{11cm}{Riccardo Argurio$^1$, Matteo Bertolini$^2$, Gabriele Ferretti$^3$,
Alberto Lerda$^4$ and  Christoffer Petersson$^3$}
\\
~\\
~\\
$^1$Physique Th\'eorique et Math\'ematique and
International Solvay Institutes \\ Universit\'e Libre de
Bruxelles,
CP 231, 1050 Bruxelles, Belgium\\\vspace{0.3cm}
$^2$SISSA/ISAS and INFN - Sezione di Trieste\\ Via
Beirut 2; I-34014 Trieste, Italy\\\vspace{0.3cm}
$^3$Department of Fundamental Physics\\ Chalmers
University of Technology, 412 96 G\"oteborg, Sweden\\\vspace{0.3cm}
$^4$Dipartimento di Scienze e Tecnologie Avanzate \\
Universit\`a del Piemonte Orientale,
I-15100 Alessandria, Italy\\
Istituto Nazionale di Fisica Nucleare  - sezione di Torino,
Italy\\
\vspace{0.3cm}

\email{rargurio@ulb.ac.be, bertmat@sissa.it, ferretti@chalmers.se, lerda@to.infn.it, chrpet@chalmers.se}\\}

\abstract{We study the effects produced
by D-brane instantons on the holomorphic quantities of a D-brane gauge theory
at an orbifold singularity.
These effects are not limited to reproducing the well known contributions
of the gauge theory instantons but also generate extra terms
in the superpotential or the prepotential.
On these brane instantons there are some neutral fermionic zero-modes
in addition to the ones expected from broken supertranslations.
They are crucial in correctly reproducing effects which
are dual to gauge theory instantons, but they may make some other interesting
contributions vanish. We analyze how orientifold projections can remove
these zero-modes and thus allow for new superpotential terms.
These terms contribute to the dynamics of the effective gauge theory,
for instance in the stabilization of runaway directions.}
\preprint{SISSA-16/2007/EP}
\begin{document}
\setcounter{section}{0}

\renewcommand{\thefootnote}{\arabic{footnote}}
\setcounter{footnote}{0} \setcounter{page}{1}

\tableofcontents

\section{Introduction}

It has long been realized that instantons in string theory are often
in close correspondence with instantons in gauge
theories~\cite{Witten:1995gx,Douglas:1995bn,Witten:1996bn,Ganor:1996pe,Green:2000ke,Billo:2002hm}.
Recently
it was found that in some situations stringy instantons can dynamically generate
some terms which from a low-energy effective point of view
enter as ordinary external couplings in the superpotential of gauge theories living on space-filling branes 
\cite{Blumenhagen:2006xt,Ibanez:2006da,Florea:2006si,Abel:2006yk,Akerblom:2006hx,Bianchi:2007fx,Cvetic:2007ku,Bianchi:fu}.
By instantons in string theory we generally mean instantons which are
geometrically realized as Euclidean extended objects wrapped on some
non-trivial cycles of the geometry. Thus, in a sense, a stringy
instanton has a ``life of its own'', not requiring an underlying
gauge theory. This opens up the possibility of having contributions originating
from instantons that do not admit a standard
gauge theory realization. We shall refer to these instantons as
{\it exotic}.

There has been some debate in the recent literature about the instances where
such exotic instantons can actually contribute to the gauge theory superpotential 
in a non-trivial manner.
In this work we will contribute to such a debate by considering backgrounds where a
simple CFT description is possible, such as orbifolds or orientifolds thereof.

We present various simple examples of what we believe to be
a rather generic situation. Namely, the presence of extra zero-modes
for these instantons, in addition to those required by the counting of
broken symmetries, makes some of their contributions vanish.
Such extra zero-modes should not come as a surprise,
since a D-brane instanton in a CY manifold
breaks a total of four out of eight supercharges, i.e.
it has two extra
fermionic zero-modes from the point of view of holomorphic $\Ncal=1$ gauge theory quantities.
We give some arguments as to why the backreaction of the space-filling branes on the geometry might not help 
in lifting these extra zero-modes. We further argue that only more
radical changes of the background, such as the introduction of fluxes,
deformations of the CY geometry or the introduction of orientifold planes,
can remove these zero-modes. When this happens, exotic instantons do
contribute to the gauge theory superpotential and may provide qualitative
changes in the low energy effective dynamics, as for instance the
stabilization of otherwise runaway directions.

We will be interested in Euclidean D-branes in type II
theories. We will work with IIB  fractional branes at orbifold and
orientifold singularities rather than type IIA wrapped branes. The
motivation for this choice of setting is two-fold. First,
recent advances in the gauge/gravity correspondence require the study
of exotic instantons, whose effects tend to stabilize the gauge
theory rather than unstabilize
it~\cite{Intriligator:2005aw,Argurio:2006ew,Florea:2006si,Argurio:2006ny}, and the
gauge/gravity correspondence is more naturally defined in the context of IIB theory.
Second, similar effects are used
in string phenomenology to try to understand possible mechanisms for
neutrino
masses~\cite{Blumenhagen:2006xt,Ibanez:2006da,Cvetic:2007ku}. This
latest activity is mainly done in the type IIA scenario, but we find it easier
to address some subtle issues in the IIB orbifold case.

While working in an exact string background, our considerations will
nonetheless be only local, {\it i.e.} we will not be concerned with global
issues such as tadpole cancellation that arise in proper
compactifications. This is perfectly acceptable in the context of the
gauge/gravity correspondence where the internal manifold is non-compact
but, even for string phenomenology, the results we obtain stand (locally)
when properly embedded in a consistent compactification.

The paper is organized as follows: In section~\ref{prelim} we set up
the notation and discuss some preliminary material. In
section~\ref{sectZ2Z2} we discuss our first case, namely the $\Ncal=1$
$\mathbf{Z}_2 \times \mathbf{Z}_2$ orbifold. After briefly recovering
the usual instanton generated corrections to the superpotential we
discuss the possible presence of additional exotic contributions and
find that they are not present because of the additional
zero-modes. We conclude by giving a CFT argument on why such zero-modes
are not expected to be lifted even by taking into account the backreaction of the
D-branes, unless one is willing to move out the orbifold point in the CY moduli space.
Sections~\ref{secOZ2} and~\ref{sectz3} present two separate
instances where exotic contributions are present after having removed
the extra zero-modes by orientifolding. The first is an $\Ncal=1$
orientifold, the second is an $\Ncal=2$ orientifold, displaying
corrections to the superpotential and the prepotential, respectively. We end with
some conclusions and a discussion of further developments.

\section{Preliminaries}
\label{prelim}

In this section we briefly review the generic setup in the well
understood $\Ncal=4$ situation in order to introduce the notation for
the various fields and moduli and their couplings. The more interesting
theories we will consider next will be suitable projections of the
$\Ncal=4$ theory. In fact, the exotic
cases can all be reduced to orbifolds/orientifolds of this master case
once the appropriate projections on the Chan-Paton factors are
performed.

Since we are interested in instanton physics (for
comprehensive reviews see~\cite{Dorey:2002ik} and the
recent~\cite{Bianchi:2007ft}) we will take the ten dimensional metric
to be Euclidean. We consider a system where both D3-branes and
D$(-1)$-branes (D-instantons) are present.
To be definite, we take $N$ D3's and $k$ D-instantons \footnote{These
D3/D$(-1)$ brane systems (and their orbifold projections) are very useful and
efficient in studying instanton effects from a stringy perspective even in the
presence of non-trivial closed string backgrounds, both of NS-NS
type \cite{Billo:2005fg} and of R-R type \cite{Billo:2004zq}.}.

Quite generically we can
distinguish three separate open string sectors:
\begin{itemize}
\item The gauge sector, made of those open strings with both ends on a
D3-brane. We assume the brane world-volumes are lying along the first
four coordinates $x^\mu$ and are orthogonal to the last six $x^a$. The
massless fields in this sector form an $\Ncal=4$ SYM
multiplet~\cite{Witten:1995im}. We denote the bosonic components by
$A_\mu$ and $X^a$.  Written in $\Ncal=1$ language this multiplet is
formed by a gauge superfield whose field strength is denoted by
$W_\alpha$ and three chiral superfields $\Phi^{1,2,3}$. With a slight
abuse of notation, the bosonic components of the chiral superfields
will also be denoted by $\Phi$, {\it i.e.} $\Phi^1 = X^4 + i X^5$ and so
on. In $\Ncal=2$ language we have instead a gauge superfield $\Acal$
and a hypermultiplet $H$, all in the adjoint representation. The low
energy action of these fields is a four dimensional $\Ncal=4$ gauge
theory. All these fields are $N \times N$ matrices for a gauge group
$\mathrm{SU}(N)$.
\item The neutral sector, which comprises the zero-modes of
strings with both ends on the D-instantons. It is usually  referred to as the
neutral sector because these modes do not transform under the gauge
group. The zero-modes are easily obtained by dimensionally reducing
the maximally supersymmetric gauge theory to zero dimensions. We will
use an ADHM~\cite{Atiyah:1978ri} inspired notation \cite{Green:2000ke,Billo:2002hm}.
We denote the bosonic fields as $a_\mu$ and $\chi^a$, where the
distinction between the two is made by the presence of the
D3-branes. The fermionic zero-modes are denoted by $M^{\alpha A}$ and
$\lambda_{\dot\alpha A}$, where $\alpha$ and $\dot\alpha$ denote the
(positive and negative) four dimensional chiralities and $A$ is an
$\mathrm{SU}(4)$ (fundamental or anti-fundamental) index denoting the chirality
in the transverse six dimensions. The ten dimensional chirality of
both fields is taken to be negative. In Euclidean space $M$ and
$\lambda$ must be treated as independent. When needed, we will also
introduce the triplet of auxiliary fields $D^c$, directly analogous to
the four dimensional $D$, that can be used to express the various
interactions in an easier form as we will see momentarily. All these
fields are $k \times k$ matrices where $k$ is the instanton number.
\item The charged sector, comprising the zero-modes of strings
stretching between a D3-brane and a D-instanton.  For each pair of
such branes we have two conjugate sectors distinguished by the orientation of the string.
In the NS sector, where the world-sheet fermions have opposite modding
as the bosons, we obtain a bosonic spinor $\omega_{\dot\alpha}$ in the
first four directions where the GSO projection picks out the negative
chirality. In the conjugate sector, we will get an independent bosonic
spinor $\bar\omega_{\dot\alpha}$ of the same chirality.  Similarly, in
the R sector, after the GSO projection we obtain a pair of independent
fermions (one for each conjugate sector) both in the fundamental of
$\mathrm{SU}(4)$ which we denote by $\mu^A$ and $\bar\mu^A$. These fields are
rectangular matrices $N \times k$ and $k \times N$.
\end{itemize}
The couplings of the fields in the gauge
sector give rise to a four dimensional gauge theory.
The instanton corrections to such a theory are
obtained by constructing the Lagrangian describing the interaction of
the gauge sector with the charged sector zero-modes while performing the integral over
\emph{all} zero-modes, both charged and neutral.
A crucial point to notice and which will be important later is
that while the neutral modes do not transform under the gauge group, their
presence affects the integral because of their coupling to the charged sector.

The part of the interaction involving only the instanton moduli
is well known from the ADHM construction and it is
essentially the reduction of the interacting gauge Lagrangian for
these modes in a specific limit where the Yukawa terms for $\lambda$
and the quadratic term for $D$ are scaled out
(see~\cite{Dorey:2002ik,Billo:2002hm} for details). The final form of this part of
the interaction is:
\beqs S_1 &=& \tr\Big\{-{[a_\mu,\chi^a]}^2 +
\chi^a \bar\omega_{\dot\alpha}\omega^{\dot\alpha} \chi_a
+\,\frac{i}{2} (\bar\Sigma^a)_{AB} \bar\mu^A \mu^B \chi_a -
\frac{i}{4}(\bar\Sigma^a)_{AB}  M^{\alpha A} {[\chi_a, M^B_\alpha]}
\nonumber \\
&+& i \left(\bar\mu^A \omega_{\dot\alpha} +
\bar\omega_{\dot\alpha} \mu^A + \sigma^\mu_{\beta
\dot\alpha}{[M^{\beta A}, a_\mu]}\right)\! \lambda^{\dot\alpha}_A
- i D^c\!\left( \bar\omega^{\dot \alpha}
(\tau^c)^{\dot\beta}_{\dot\alpha} \omega_{\dot\beta} + i
\bar\eta^c_{\mu\nu}  {[a^\mu, a^\nu]}\right) \!\Big\}
\label{S1}
\eeqs
where the sum over colors and instanton indices is understood. $\tau$
denotes the usual Pauli matrices, $\bar\eta$ (and $\eta$) the 't~Hooft
symbols and $\bar\Sigma$ (and $\Sigma$) are used to construct the
six-dimensional gamma-matrices
\beq
\Gamma^a = \begin{pmatrix}0 &
\Sigma^a \cr \bar\Sigma^a & 0 \cr\end{pmatrix}~.
\label{gammama}
\eeq
The above interactions can all be understood in terms of string
diagrams on a disk with open string vertex operators inserted at the
boundary in the $\alpha' \to 0$ limit.

The interaction of the charged sector with the scalars of the gauge
sector can be worked out in a similar way and yields
\beq
S_2 = \tr\Big\{
\bar\omega_{\dot\alpha } X^a X_a \omega^{\dot\alpha } + \frac{i}{2}
(\bar\Sigma^a)_{AB} \bar\mu^A X_a  \mu^B \Big\}~.
\label{S2}
\eeq
Let us rewrite the above action in a way which will be more illuminating
in the following sections. Since we will be mainly focusing on situations
where we have ${\cal N}=1$ supersymmetry, it is useful to write explicitly
all indices in $\mathrm{SU}(4)$ notation, and then break them into
$\mathrm{SU}(3)$ representations.
We thus write the six scalars $X_a$ as the antisymmetric representation of
$\mathrm{SU}(4)$ as follows
\beq
X_{AB}=-X_{BA}\equiv (\bar\Sigma^a)_{AB} X_a~.
\eeq
The action $S_2$ then reads
\beq
S_2 = \tr \Big\{
\frac18\,\epsilon^{ABCD} \bar\omega_{\dot\alpha } X_{AB} X_{CD} \omega^{\dot\alpha }
+ \frac{i}{2}\, \bar\mu^A X_{AB}  \mu^B \Big\}~.
\label{S2prime}
\eeq
Splitting now the indices $A$ into $i=1\dots 3$ and 4, we can identify
$\Phi^\dagger_i \equiv X_{i4}$ in the $\bf{\bar 3}$ of $\mathrm{SU}(3)$
and $\Phi^i \equiv \frac{1}{2} \epsilon^ {ijk}X_{jk}$ in the $\bf{3}$
of $\mathrm{SU}(3)$. Thus we can rewrite the action (\ref{S2prime}) as
\beq
S_2 = \tr \Big\{\frac12\,
\bar\omega_{\dot\alpha }\big\{ \Phi^i ,\Phi^\dagger_i \big\}\omega^{\dot\alpha }
+ \frac{i}{2} \,\bar\mu^i \Phi^\dagger_i \mu^4
- \frac{i}{2} \,\bar\mu^4 \Phi^\dagger_i \mu^i
-\frac{i}{2} \,\epsilon_{ijk} \bar\mu^i \Phi^j \mu^k \Big\}~.
\label{S2holo}
\eeq
In the above form, it is clear which zero-modes couple to the holomorphic
superfields and which others couple to the anti-holomorphic ones. This
distinction will play an important role later.

The main object of our investigation is the integral of $e^{-S_1 - S_2}$
over \emph{all} moduli
\beq
Z = \Ccal \int
d\{a,\chi,M,\lambda,D,\omega,\bar\omega,\mu,\bar\mu\} \,e^{-S_1 -S_2}~,
\label{master}
\eeq
where we have lumped all field independent normalization constants (including
the instanton classical action and the appropriate powers of $\alpha^\prime$
required by dimensional analysis) into an overall coefficient $\Ccal$.
There are, of course, other interactions involving the fermions and
the gauge bosons but, as far as the determination of the holomorphic
quantities are concerned, they can be obtained from the previous ones
and supersymmetry arguments. For example, a term in the superpotential
is written as the integral over chiral superspace $\int dx^4
d\theta^2$ of a holomorphic function of the chiral superfields, but
such a function is completely specified by its value for bosonic
arguments at $\theta=0$. Thus, if we can ``factor out'' a term $\int
dx^4 d\theta^2$ from the moduli integral (\ref{master}), whatever is
left will define the complex function to be used in the superpotential
and similarly for the prepotential in the $\Ncal = 2$ case if we
succeed in factoring out an integral over $\Ncal=2$ chiral superspace
$\int dx^4 d\theta^4$.

The coordinates $x$ and $\theta$ must of course come from the
(super)translations broken by the instanton and they will be associated to
the center of mass motion of the D-instanton,
namely, $x^\mu = \tr a^\mu$ and
$\theta^{\alpha A} = \tr M^{\alpha A}$ for some values of $A$.\footnote{Obviously, for the case of an anti-instanton, the
roles of $M$ and $\lambda$ are reversed.} One
must pay attention however to the presence of possible additional
neutral zero-modes coming either from the traceless parts of the above
moduli or from the fields $\lambda$ and $\chi$. These modes must also
be integrated over in (\ref{master}) and their effects, as we shall
see, can be quite dramatic. In particular, the presence of $\lambda$
in some instances is crucial for the implementation of the usual ADHM
fermionic constraints whereas in other circumstances it makes the whole
contribution to the superpotential vanish. These extra $\lambda$
zero-modes are ubiquitous in orbifold theories and generically make it difficult
to obtain exotic instanton corrections for these models. As we shall
see, they can however be easily projected out by an orientifold construction
making the derivation of such terms possible.

In the full expression for the instanton corrections there will also
be a field-in\-de\-pen\-dent normalization factor coming from the
one-loop string diagrams and giving for instance the proper $g_{YM}$
dependence in the case of the usual instanton corrections. In this
paper we will only focus on the integral over the zero-modes, which
gives the proper field-dependence, referring the reader to
\cite{Abel:2006yk,Akerblom:2006hx} for a discussion of these other
issues.

\section{The $\Ncal = 1~ \mathbf{Z}_2 \times \mathbf{Z}_2$ orbifold}
\label{sectZ2Z2}

In order to present a concrete example of the above discussion, let us
study a simple $\mathbf{C}^3/\mathbf{Z}_2 \times \mathbf{Z}_2$
orbifold singularity. The resulting $\Ncal=1$ theory is a non-chiral
four-node quiver gauge theory with matter in the bi-fundamental. Non-chirality implies that
the four gauge group ranks can be chosen independently~\cite{Bertolini:2001gg}. This
corresponds to being able to find a basis of three independent fractional branes in
the geometry (for a review on fractional branes on orbifolds see e.g.~\cite{Bertolini:2003iv}).

\begin{figure}[ht]
\begin{center}
{\includegraphics{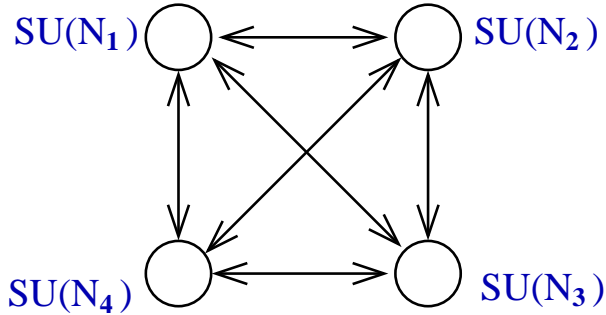}}
\caption{\small  Quiver diagram for the
$\mathbf{Z}_2 \times \mathbf{Z}_2$ orbifold theory. Round circles
correspond to $\mathrm{SU}(N_\ell)$ gauge factors while the lines connecting
quiver nodes represent the bi-fundamental chiral superfields
$\Phi_{\ell m}$.}
\label{quiverD3}
\end{center}
\end{figure}

The field content can be conveniently summarized in a quiver diagram,
see Fig.~\ref{quiverD3}, which, together with the cubic superpotential
\beqs  W &=&
\Phi_{12}\Phi_{23}\Phi_{31}-\Phi_{13}\Phi_{32}\Phi_{21}+\Phi_{13}\Phi_{34}\Phi_{41}-
\Phi_{14}\Phi_{43}\Phi_{31} \nonumber \\ &&
+\Phi_{14}\Phi_{42}\Phi_{21}-\Phi_{12}\Phi_{24}\Phi_{41}+\Phi_{24}\Phi_{43}\Phi_{32}-
\Phi_{23}\Phi_{34}\Phi_{42}~,
         \label{wtree}
\eeqs 
uniquely specifies the theory.

A stack of $N$ regular D3-branes amounts to having one and the same
rank assignment on the quiver. The gauge group is then $\mathrm{SU}(N)^4$ and the theory
is an $\Ncal =1$ SCFT. Fractional branes correspond instead to different
(but anomlay free) rank assignments. Quite generically,
fractional branes can be divided into three different classes,
depending on the IR dynamics they trigger~\cite{Franco:2005zu}. The
non-chiral nature and the particularly symmetric structure of the
orbifold under consideration allows one to easily construct any
such instance of fractional brane class.

If we turn on a single node, we are left with a pure $\mathrm{SU}(N)$ SYM gauge
theory, with no matter fields and no superpotential. This theory is
believed to confine. The geometric dual effect is that the
corresponding fractional brane leads to a geometric transition where
the branes disappear leaving behind a deformed geometry. Indeed, there
is one such deformation in the above singularity.

Turning on two nodes leads already to more varied phenomena. There are
now two bi-fundamental superfields, but still no tree level
superpotential. Thus, the system is just like two coupled massless
SQCD theories or, by a slightly asymmetric point of view, massless
SQCD with a gauged diagonal flavor group. The low-energy behavior
depends on the relative ranks of the two nodes.

If the ranks are different, the node with the highest rank is in a
situation where it has less flavors than colors. Then an
Affleck-Dine-Seiberg (ADS)
superpotential~\cite{Taylor:1982bp,Affleck:1983mk} should be
dynamically generated, leading eventually to a runaway behavior. This
set up of fractional branes is sometimes referred to as
supersymmetry breaking fractional branes \cite{Berenstein:2005xa,Franco:2005zu,Bertolini:2005di}.

If the ranks are the same we are in a situation similar to $N_f=N_c$
SQCD for both nodes. Hence we expect to have a moduli space of SUSY vacua,
which gets deformed, but not lifted, at the quantum level. This moduli space is
roughly identified in the geometry with the fact that the relevant fractional branes
are interpreted as D5-branes wrapped on the 2-cycle of a
singularity which is locally
$\mathbf{C}\times (\mathbf{C}^2/\mathbf{Z}_2)$. Such a fractional
brane can move in the $\mathbf{C}$ direction. This is what is called
an ${\cal N}=2$ fractional brane since, at least geometrically, it
resembles very much the situation of fractional branes at
${\cal N}=2$ singularities.

In what follows we use the two-node example as a simple setting in which we
can analyze the subtleties involved in the integration over the neutral modes.
For the gauge theory instanton case it is known that there are {\it extra}
neutral fermionic zero-modes in addition to those required to generate the
superpotential. Their integration allows to recover the fermionic ADHM
constraints on the moduli space of the usual field theory instantons.
For such instantons, we will be able to obtain the ADS superpotential and corresponding runaway
behavior in the familiar context with
$N_c$ and $N_f$ fractional branes at the respective nodes, for $N_f = N_c -1$.
On the other hand, we will argue that the presence of such extra zero-modes
rules out the possibility of having exotic instanton effects, such as terms involving
baryonic operators in the $N_f=N_c$ case.
It was the desire to study such possible
contributions that constituted the original motivation for this investigation.
We will first show that such effects are absent for this theory as it stands, and we will
later discuss when and how this problem can be cured.\footnote{In a situation where the CFT description is less under control than in
the setting discussed in the present paper,
it has been argued in \cite{Argurio:2006ny} that such baryonic couplings do arise
in the context of fractional branes on orbifolds of the conifold, possibly
at the expense of introducing O-planes. Also in a IIA set up similar to
the ones of \cite{Blumenhagen:2006xt,Ibanez:2006da,Abel:2006yk,Akerblom:2006hx,Cvetic:2007ku} 
it seems reasonable that one can wrap an
ED2-brane along an O6-plane and produce such couplings on other
intersecting D6-branes.}

Our orbifold theory can be easily obtained as an orbifold projection of $\Ncal
= 4$ SYM. The orbifolding procedure and the derivation of the
superpotential (\ref{wtree}) are by now standard. We briefly recall
the main points in order to fix the notation and because some of the
details will be useful later in describing the instantons in such a
set up.

The group $\mathbf{Z}_2 \times \mathbf{Z}_2$ has four elements: the
identity $e$, the generators of the two $\mathbf{Z}_2$ that we denote
with $g_1$ and $g_2$ and their product, denoted by $g_3 =g_1 g_2$. If
we introduce complex coordinates $(z_1, z_2,z_3)\in \mathbf{C}^3$
\beq
z^1 = x^4 + i x^5~~,~~z^2 = x^6 +i x^7~~,~~ z^3 = x^8 + i x^9
\label{zzz67}
\end{equation}
the action of the orbifold group can be defined as in Table~1.

\begin{table} [ht]
\begin{center}
\begin{tabular}{c||c|c|c|}
&$z^1$ & $z^2$ & $z^3$ \\ \hline  \hline $e$    &$z^1$ & $z^2$ & $z^3$
\\  \hline $g_1$  &$z^1$ & $-z^2$ & $-z^3$ \\  \hline $g_2$ &$-z^1$ &
$z^2$ & $-z^3$ \\  \hline $g_3$  &$-z^1$ & $-z^2$ & $z^3$ \\ \hline
\end{tabular}
\caption{\small The action of the orbifold generators.}
\end{center}
\end{table}

Let $\gamma(g)$ be the regular representation of the orbifold group on
the Chan-Paton factors. If the orbifold is abelian, as always in the
cases we shall be interested in, we can always diagonalize all
matrices $\gamma(g)$. We will assume that the two generators have the
following matrix representation
\beq
\gamma(g_1) =\sigma_3 \otimes
\mathbf{1} =\begin{pmatrix}1 & 0 & 0   & 0   \cr  0 & 1 & 0   & 0 \cr 0 & 0
& -1   &  0 \cr  0   & 0   & 0 & -1 \cr\end{pmatrix}~~~,~~~
\gamma(g_2) =\mathbf{1}\otimes
\sigma_3=\begin{pmatrix}1 & 0 & 0   & 0   \cr   0 & -1 & 0   & 0 \cr 0 & 0
& 1   &  0 \cr  0   & 0   & 0 & -1 \cr\end{pmatrix}~
\label{chanpatonz2z2}
\eeq
where the 1's denote $N_\ell \times N_\ell$ unit matrices ($\ell=1,...,4$).
Then, the orbifold projection amounts to enforcing the conditions
\beq
A_\mu = \gamma(g)A_\mu \gamma(g)^{-1}~~~,~~~
\Phi^i = \pm \gamma(g)\Phi^i\gamma(g)^{-1}
\label{orbaction}
\eeq
where the sign $\pm$ must be chosen according to the action of the orbifold generators $g$
that can be read off from Table~1.
With the choice (\ref{chanpatonz2z2}), the vector superfields
are block diagonal matrices of different size $(N_1,N_2,N_3,N_4)$, one for
each node of the quiver, while the three chiral superfields $\Phi^i$
have the following form~\cite{Bertolini:2001gg}
\beq
\Phi^1 = \begin{pmatrix}0 & \times &
0   & 0   \cr  \times & 0 & 0   & 0 \cr 0 & 0   & 0   &  \times \cr  0
& 0   & \times & 0 \cr\end{pmatrix},~~ \Phi^2 = \begin{pmatrix}0 & 0
& \times &  0  \cr     0   & 0   & 0   & \times \cr \times & 0   & 0
& 0  \cr  0   & \times & 0 & 0 \cr\end{pmatrix}, ~~ \Phi^3 =
\begin{pmatrix}0 & 0   & 0   & \times \cr  0   & 0 & \times & 0   \cr
0 & \times & 0   & 0   \cr  \times & 0   & 0   & 0 \cr\end{pmatrix}~,
\label{structure}
\eeq
where the crosses represent the non-zero entries $\Phi_{\ell m}$ appearing
in the superpotential (\ref{wtree}).

\subsection{Instanton sector}

Now consider D-instantons in the above set up. Such instantons
preserve half of the 4 supercharges preserved by the system of
D3-branes plus orbifold. In this respect recall that the fractional
branes preserve exactly the same supercharges as the regular
branes.\footnote{There is another Euclidean brane which preserves two
supercharges, namely the Euclidean (anti) D3-branes orthogonal to the
4 dimensions of space-time. We will be considering here only the
D-instantons, leaving the complete analysis of the other effects to
future work. In this context, note that the extended brane instantons
would have an infinite action (and thus a vanishing contribution) in
the strict non-compact set up we are using here.}
Using the $\Ncal=4$ construction of the previous section and the
structure of the orbifold presented in eq.~(\ref{structure}), we
now proceed in describing the zero-modes for such instantons.

The neutral sector is very similar to the gauge sector.
Indeed, in the $(-1)$~superghost picture, the vertex
operators for such strings will be exactly the same, except for the
$e^{ip\cdot X}$ factor which is absent for the instanton. The
Chan-Paton structure will also be the same, so that the same pattern
of fractional D-instantons will arise as for the fractional
D3-branes. In particular, the only regular D-instanton (which could be
thought of as deriving from the one of $\Ncal=4$ SYM) is the one with
rank (instanton number) one at every node. All other situations can be
thought of as fractional D-instantons, which can be interpreted as Euclidean D1-branes wrapped
on the two-cycles at the singularity, ED1 for short. Generically, we can
then characterize an instanton configuration in our orbifold by $(k_1, k_2,
k_3, k_4)$.

Following the notation introduced in section~\ref{prelim}, the bosonic
modes will comprise a $4\times 4$ block diagonal matrix $a^\mu$, and
six more matrix fields $\chi^1, \dots \chi^6$, that can be paired into
three complex matrix fields $\chi^1 + i\chi^2, \chi^3 + i\chi^4,
\chi^5 + i\chi^6$, having the same structure as (\ref{structure}) but
now where each block entry is a $k_\ell\times k_m$ matrix. On the fermionic
zero-modes  $M^{\alpha A}$ and $\lambda_{\dot\alpha A}$ (also
matrices) the orbifold projection enforces the conditions
\beq
M^{\alpha A} = R(g)^A_{~ B}\,\gamma(g) M^{\alpha B} \gamma(g)^{-1}
~~~,~~~
\lambda_{\dot\alpha A} =\,\gamma(g) \lambda_{\dot\alpha B} \gamma(g)^{-1} R(g)_{~A}^B
\label{orbactionferm}
\eeq
where $R(g)$ is the orbifold action of Table~1
in the spinor representation which can be chosen
as
\beq
R(g_1) = - \Gamma^{6789}~~~,~~~
R(g_2) = - \Gamma^{4589}~.
\label{rg}
\eeq
It is easy to find an explicit representation of the Dirac matrices such that
$M^{\alpha A}$ and $\lambda_{\dot\alpha A}$ for $A=1,2,3$ also have
the structure of (\ref{structure}) while for $A=4$ they are block diagonal.
Equivalently, one could write the spinor indices in the internal space in terms
of the three $\mathrm{SO}(2)$~charges associated to the embedding $\mathrm{SO}(2)\times
\mathrm{SO}(2)\times
\mathrm{SO}(2) \subset \mathrm{SO}(6) \simeq \mathrm{SU}(4)$
\beqs
&& M^{\alpha-++} = M^{\alpha 1}~,~
M^{\alpha+-+} = M^{\alpha 2}~,~
M^{\alpha++-} = M^{\alpha 3}~,~
M^{\alpha---} = M^{\alpha 4}~,~ \nonumber \\
&& \lambda_{\dot\alpha+--}
= \lambda_{\dot\alpha 1}~,~
\lambda_{\dot\alpha-+-}
= \lambda_{\dot\alpha 2}~,~
\lambda_{\dot\alpha--+}
= \lambda_{\dot\alpha 3}~,~
\lambda_{\dot\alpha+++}
= \lambda_{\dot\alpha 4}~.
\eeqs
The most notable difference between the neutral sector and the
gauge theory on the D3-branes is that, whereas in the four-dimensional
theory the $U(1)$ gauge factors are rendered massive by a
generalization of the Green-Schwarz mechanism and do not appear in the
low energy action, for the instanton they are in fact present and
enter crucially into the dynamics.

Let us finally turn to the charged sector, describing strings going
from the instantons to the D3-branes. The analysis of the spectrum and
the action of the orbifold group on the Chan-Paton factors show, in particular, 
that the bosonic zero-modes are diagonal in the gauge factors. There are four block 
diagonal matrices of bosonic zero-modes $\omega_{\dot\alpha}, ~\bar\omega_{\dot\alpha}$ 
with entries $N_\ell \times k_\ell$ and $k_\ell \times N_\ell$ respectively and 
eight fermionic matrices $\mu^A, ~\bar\mu^A$  with entries $N_\ell
\times k_m$ and $k_m \times N_\ell$, that again display the same
structure as above -- same as (\ref{structure}) for $A=1,2,3$ and
diagonal for $A=4$.

\subsection{Recovery of the ADS superpotential}
\label{recoverADS}
The measure on the moduli space of the instantons and the ADHM
constraints are simply obtained by inserting the above expressions
into the moduli integral (\ref{master}). If one chooses some of the
$N_\ell$ or $k_\ell$ to vanish one can deduce immediately from the structure
of the projection which modes will survive and which will not.

As a consistency check, one can try to reproduce the ADS correction to
the superpotential~\cite{Taylor:1982bp,Affleck:1983mk} for the theory
with two nodes. Take fractional branes corresponding to a rank
assignment $(N_c,N_f,0,0)$, and consider the effect of a ED1 corresponding
to instanton numbers $(1,0,0,0)$.

The only chiral fields present are the two components of $\Phi^1$
connecting the first and second node
\beq
\Phi^1 =
\begin{pmatrix}0 & Q & 0   & 0   \cr  \tilde Q & 0 & 0   & 0 \cr 0 & 0
& 0   &  0 \cr  0   & 0   & 0 & 0 \cr\end{pmatrix}~.
\eeq
Since the instanton is sitting only at one node, all off diagonal
neutral modes are absent, as they connect instantons at two
distinct nodes. Thus, the only massless modes present in the neutral
sector are four bosons $x^\mu$, denoting the upper-left component of
$a^\mu$, two fermions $\theta^\alpha$ denoting the upper-left
component of $M^{\alpha 4}$ and two more fermions
$\lambda_{\dot\alpha}$ denoting the upper-left component of
$\lambda_{\dot\alpha 4}$. We have identified
the non zero entries of $a^\mu$ and $M^{\alpha 4}$ with the
super-coordinates  $x^\mu$ and $\theta^\alpha$ since they precisely
correspond to the Goldstone modes of the super-translation symmetries broken by the
instanton and do not appear in $S_1 + S_2$ (cfr. (\ref{S1}) and
(\ref{S2})). Their integration produces the integral over space-time
and half of Grassmann space which precedes the superpotential term to
which the instanton contributes. On the contrary,
$\lambda_{\dot\alpha}$ appears in $S_1$ and when it is integrated
it yields the fermionic ADHM constraint.

In the charged sector, we have bosonic zero-modes
$\omega_{\dot\alpha}^u$ and $\bar\omega_{{\dot\alpha}u}$, with $u$ an
index in the fundamental or anti-fundamental of $\mathrm{SU}(N_c)$. In
addition, there are fermionic zero-modes $\mu^u$ and $\bar\mu_u$ with
indices in $\mathrm{SU}(N_c)$, together with additional fermionic zero-modes
$\mu'{}^f$ and $\bar\mu'_f$ where the index $f$ is now in the
fundamental or anti-fundamental of $\mathrm{SU}(N_f)$.\footnote{Recall that the bosonic zero-modes are diagonal in the gauge
 factors; therefore there are no $\omega_{\dot\alpha}^f$ and
 $\bar\omega_{{\dot\alpha}f}$ zero-modes.} Note that the $\mu$
zero-modes carry an $\mathrm{SU}(4)$ index 4 (being on the diagonal) while the $\mu'$
zero-modes carry an $\mathrm{SU}(4)$ index $1$, since they are of the same form as
$\Phi^1$.

\begin{figure}[ht]
\begin{center}
{\includegraphics{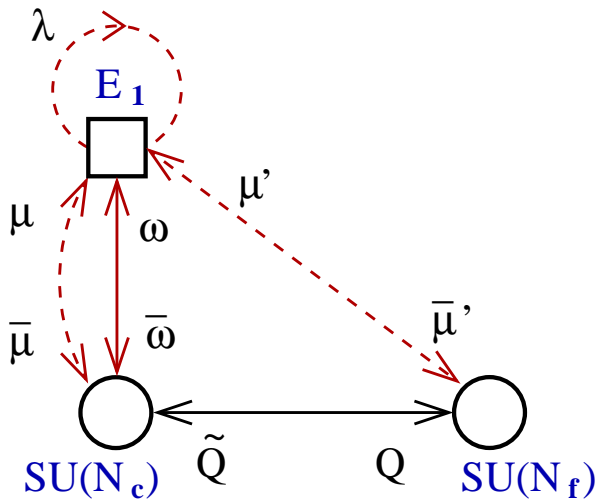}}
\caption{\small  Quiver diagram describing an ordinary instanton
in a $\mathrm{SU}(N_c) \times \mathrm{SU}(N_f)$ theory. Gauge theory nodes are represented by
round circles, instanton nodes by squares. The ED1 is wrapped on the same cycle
as the color branes. All zero-modes are included except the $\theta$'s and the $x^\mu$'s,
which only contribute to the measure for the integral over chiral superspace.}
\label{quiverADS}
\end{center}
\end{figure}

All this can be conveniently summarized in a generalized quiver diagram as represented in
Fig.~\ref{quiverADS}, which accounts for both the brane configuration and the
instanton zero-modes.

For a single instanton, the action (\ref{S1}) greatly simplifies since
many fields are vanishing as well as all commutators and one gets
\beq
S_1 = i
\left(\bar\mu_u \omega^u_{\dot\alpha} +  \bar\omega_{\dot\alpha u}
\mu^u \right) \lambda^{\dot\alpha}  - i D^c \bar\omega_{u}^{\dot
\alpha}(\tau^c)_{\dot\alpha}^{\dot\beta} \omega_{\dot\beta}^{u}~.
\label{S1ADS}
\eeq
Similarly, the coupling of the charged modes to the chiral
superfield can be expressed by writing eq. (\ref{S2}) as
\beq
S_2 =
\frac12\,\bar\omega_{{\dot\alpha}u} \big(Q^u_f {Q^\dagger}{}^f_v
+\tilde Q^\dagger{}^u_f \tilde Q^f_v\big) \omega^{{\dot\alpha}v} -
\frac{i}2\,\bar \mu_u \tilde Q^\dagger{}^u_f \mu'{}^f +\frac{i}2\,
\bar \mu'_f Q^\dagger{}^f_u \mu^u~.
          \label{mixed}
\eeq
Note that it is the anti-holomorphic superfields that enter in the couplings
with the fermionic zero-modes, as is clear by comparing with
(\ref{S2holo}). The above action is exactly the same which appears
in the ADHM construction as reviewed in \cite{Dorey:2002ik}.

We are now ready to perform the integral (\ref{master}) over all the
existing zero-modes. Writing
\beq
Z = \int dx^4 d\theta^2 \,W~,
\eeq
we see that the instanton induced superpotential is
\beq
W = \Ccal \int d\{\lambda,D,\omega,\bar\omega,\mu,\bar\mu\}\, e^{-S_1
-S_2}~.
\eeq
The integrals over $D$ and $\lambda$ enforce the bosonic
and fermionic ADHM constraints, respectively. Thus
\beq
W = \Ccal \int
d\{\omega,\bar\omega,\mu,\bar\mu\}\, \delta(\bar\mu_u
\omega^u_{\dot\alpha} +  \bar\omega_{\dot\alpha u}
\mu^u)\,\delta(\bar\omega_{u}^{\dot
\alpha}(\tau^c)_{\dot\alpha}^{\dot\beta}  \omega_{\dot\beta}^{u})
\,e^{-S_2}~.
\eeq
We essentially arrive at the point of having to
evaluate an integral over a set of zero-modes which is exactly the
same as the one discussed in detail in the literature,
{\it e.g.}~\cite{Dorey:2002ik}. We thus quickly go to the
result referring the reader to the above review for further
details. First of all, it is easy to see that, due to the presence of
extra $\mu$ modes in the integrand from the fermionic delta function,
only when $N_f=N_c-1$ we obtain a non-vanishing result. After having
integrated over the $\mu$ and $\mu'$, we are left with a
(constrained) gaussian integration that can be performed {\it e.g.} by going
to a region of the moduli space where the chiral fields are
diagonal, up to a row/column of zeroes.
Furthermore, the D-terms in the gauge sector constrain the quark
superfields to obey $Q Q^\dagger= \tilde Q^\dagger \tilde Q$, so that
the bosonic integration brings the square of a simple determinant in the
denominator.
The last fermionic integration
conspires to cancel the anti-holomorphic contributions and gives
\beq
W_{ADS} =  \frac{\Lambda^{2N_c +1}}{\det
(\tilde Q Q) }~,
\label{ADS}
\eeq
which is just the expected ADS superpotential for $N_f =N_c-1$,
the only case where such non-perturbative contribution is generated by a
genuine one-instanton effect and not by gaugino condensation. In (\ref{ADS})
$\Lambda$ is the SQCD strong
coupling scale that is reconstructed by the combination of $e^{-8\pi^2/g^2}$ coming
from the instanton action with various dimensional factors coming from the normalization
of the instanton measure \cite{Dorey:2002ik}.

\subsection{Absence of exotic contributions}

Until now, we have reproduced from stringy considerations the effect
that is supposed to be generated also by instantons in the gauge
theory.  Considering a slightly different set up, we would like to
study the possibility of generating other terms.

\begin{figure}[ht]
\begin{center}
{\includegraphics{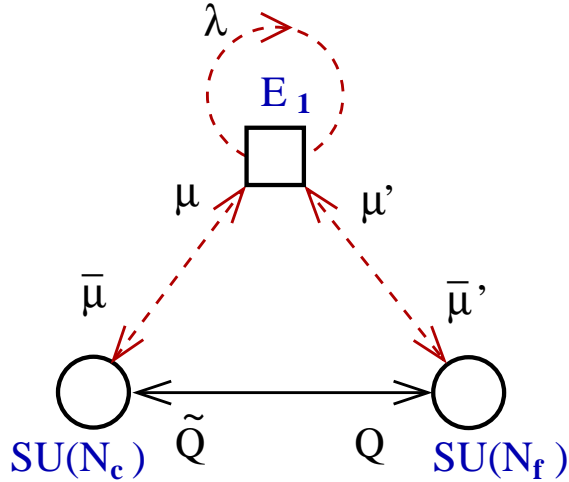}}
\caption{\small  Quiver diagram describing an exotic instanton
in a $\mathrm{SU}(N_c) \times \mathrm{SU}(N_f)$ theory. Gauge theory nodes are represented by
round circles, instanton nodes by squares. The ED1 is wrapped on a different
cycle with respect to both sets of quiver branes.}
\label{quiverBB}
\end{center}
\end{figure}

Let us consider a system with rank assignment
$(N_c,N_f,0,0)$, as before, but fractional instanton numbers
$(0,0,1,0)$. In other words, we study the effect of a single
fractional instanton sitting on an {\it unoccupied} node of the gauge
theory. The quiver diagram, with the relevant zero-modes structure,
is given in Fig.~\ref{quiverBB}.

The neutral zero-modes of the instanton sector are the same as
before. This is because the quantization of this sector does not know
the whereabouts of the D3-branes and thus all nodes are equivalent, in this respect. In
the mixed sector, we have no bosonic zero-modes now, since
the $\omega$ and $\bar\omega$ are diagonal. Note that, although we always have four mixed (ND)
boundary conditions, due to the quiver structure induced by the orbifold, here we effectively realize the same
situation one has when there are eight ND directions, namely that the bosonic sector of the charged moduli
is empty.

On the other hand, there
are fermionic zero-modes $\mu^u$, $\bar\mu_u$, $\mu'{}^f$ and
$\bar\mu'_f$, as in the previous case. Note that despite having the same
name, these zero-modes correspond actually to different Chan-Paton
matrix elements with respect to the previous ones, the difference
being in the instanton index that is not written explicitly. In particular
we can think of $\mu$ and $\mu'$ as carrying an $\mathrm{SU}(4)$ index 2 and
3 respectively.

Because of the absence of bosonic charged modes, the action (\ref{S1})
is identically zero and the action (\ref{S2}) contains only the last
term:
\beqs
S_1 &=& 0 \nonumber \\ S_2 &=& \frac{i}{2}\,\bar \mu_u Q^u_f \mu'{}^f - \frac{i}{2}\,
\bar\mu'_f \tilde Q^f_u \mu^u.
\eeqs
Note that in this case it is the holomorphic superfields which appear
above, as is clear from (\ref{S2holo}) and from noticing that
the diagonal fermionic zero-mode $\mu^4$ is not present.
We are thus led to consider
\beq
\label{exotic}
W = \Ccal \int d\{\lambda,D,\mu,\bar\mu\}\, e^{-S_2}~.
\eeq
One notices right away that the integral over the charged modes is non
vanishing (only) for the case $N_f = N_c$ and gives a tantalizing
contribution proportional to $B \tilde B$, where $B=\det Q$ and
$\tilde B=\det \tilde Q$ are the baryon fields of the theory.
However, we must carefully analyze the
integration over the remaining zero-modes of the neutral sector. Now
neither $D$ nor $\lambda$ appear in the integrand. The integral over
$D$ does not raise any concern: it is, after all, an auxiliary field
and its disappearance from the integrand is due to the peculiarities
of the ADHM limit. Before taking this limit, $D$ appeared
quadratically in the action and could be integrated out, leaving an
overall normalization constant. The integral over $\lambda$ is another
issue. In this case, $\lambda$ is absent from the integrand even
before taking the ADHM limit and its integration multiplies the above
result by zero, making the overall contribution of such instantons to
the superpotential vanishing. Of course, the presence of such extra zero-modes
should not come as a surprise since they correspond to the two extra broken
supersymmetries of an instanton on a CY.

Therefore we see that the neutral zero-modes contribution, in the exotic
instanton case, plays a dramatic role and conspires to make everything
vanishing (as opposite to the ADS case analyzed before).
A natural question is to see whether these zero-modes get
lifted by some effect we have not taken into account, yet. For one thing,
supersymmetry arguments would make
one think that taking into account the back-reaction of the D3-branes might change things.
However, in the following subsection we show that this seems not
to be the case.

\subsection{Study of the back-reaction}

Let us stick to the case $N_f=N_c$, which is the only one where the integral
(\ref{exotic}) might give a non-vanishing contribution. In this case the fractional
brane system is nothing but a stack
of ($N_c$) $\Ncal = 2$ fractional branes. These branes couple to
only one of the 3 closed string twisted sectors \cite{Bertolini:2001gg}.
More specifically, they source the metric $h_{\mu\nu}$, the R-R four-form potential
$C_{\mu\nu\rho\sigma}$ and two twisted scalars $b$ and $c$ from the
NS-NS and R-R sector respectively. This means that the disk one-point
function of their vertex operators~\cite{Friedan:1985ge,Dixon:1986qv}
is non vanishing
when the disk boundary is attached to such D3-branes. (Indeed in this way
or, equivalently, by using the boundary-state
formalism~\cite{DiVecchia:1997pr,DiVecchia:1999rh},
one can derive the profile for these fields.)

If the back-reaction of these fields on the instanton lifted the extra zero-modes $\lambda$'s,
this should be visible when computing the one point function
of the corresponding closed string vertex
operators on a disk with insertions on this boundary of the vertex operators for such moduli.
To see whether such coupling is there, we first need to write down the vertex
operators for the $\lambda$'s in the $(\pm 1/2)$~superghost pictures. The vertex in the
$(-1/2)$ picture is found {\it e.g.} in~\cite{Billo:2002hm} and reads
\beq
V^{-1/2}_\lambda(z) = \lambda_{\dot\alpha A} S^{\dot\alpha}(z) S^A(z)
e^{-\phi(z)/2}~,
\label{minushalf}
\eeq
where $S^{\dot\alpha}(z)$ and $S^A(z)$ are the spin-fields in the first four and
last six directions respectively. For our argument we need to focus on the $S^A(z)$
dependence. Since the modulus that survives the orbifold projection
is, with our conventions, $\lambda_{\dot\alpha 4} =
\lambda_{\dot\alpha +++}$, we write the corresponding spin-field as
\beq
S^{+++}(z) = e^{i H_1(z)/2} e^{i H_2(z)/2} e^{i H_3(z)/2},
\label{spinfield}
\eeq
where $H_i(z)$ is the free boson used to bosonize the fermionic sector in the $i$-th
complex direction: $\psi^i(z) = e^{i H_i(z)}$. The vertex operator in the $+1/2$~picture
can be obtained by applying the picture-changing operator
to~(\ref{minushalf})
\beq
V^{1/2}_\lambda(z) = {[Q_\mathrm{BRST}, \xi
V^{-1/2}_\lambda(z)]}~.
\eeq
The crucial part in $Q_\mathrm{BRST}$
is~\cite{Friedan:1985ge}
\beq
Q_\mathrm{BRST} = \oint \frac{dz}{2 \pi
i} \,\, \eta\,e^{\phi} \left( \psi^\mu \partial X^\mu + \bar\psi^i
\partial Z^i + \psi^i \partial \bar Z^i\right)  + \dots
\label{qrelevant}
\eeq
Because of the nature of the supercurrent, we
see that (\ref{qrelevant}) flips at most one sign in
(\ref{spinfield}), hence the product $V^{-1/2}_\lambda
V^{1/2}_\lambda$ will always carry an unbalanced charge in some of the
three internal $\mathrm{SO}(2)$ groups. On the other hand, the vertex operators for the
fields sourced by the fractional D3's cannot compensate such an unbalance. Hence,
their correlation function on the D-instanton with the insertion of
$V^{-1/2}_\lambda V^{1/2}_\lambda$ carries a charge unbalance and therefore
vanishes. Therefore, at least within the above perturbative approach,
the neutral zero-modes seem not to get lifted by the back-reaction of the D3-branes.

One might consider some additional ingredients which could provide the
lifting. A natural guess would be
moving in the CY moduli space or adding suitable background fluxes 
\cite{Martucci:2005rb,Bergshoeff:2005yp}. There are indeed non-vanishing background fields at the orbifold point, 
{\it i.e.} the $b$ fields of the twisted sectors which the $\Ncal =2$ fractional branes do not couple to.
These fields, however, being not associated to geometric deformations of the internal
space should be described by a CFT vertex operator uncharged under the $\mathrm{SO}(2)$'s,
simply because of Lorentz invariance in the internal space.
Therefore, the only way to get an effective mass term for the zero-modes $\lambda$
would be to move out of the orbifold point in the
CY moduli space. Indeed, the other moduli of the NS-NS twisted
sector, being associated to geometric blow-ups of the
singularity, are charged under (some of) the
internal $\mathrm{SO}(2)$'s and can have a non vanishing coupling with
the $\lambda$'s. More generically, complicated
closed string background fluxes might be suitable. This is an interesting
option which however we do not pursue here, since we want to stick to
situations where a CFT description is available.

A more radical thing to do is to remove the zero-modes from the very start,
for instance by means of an orientifold projection ~\cite{Pradisi:1988xd,Gimon:1996rq}.
This is the option we are
going to consider in the remainder of this work.

\section{The $\Ncal = 1$ $\mathbf{Z}_2 \times \mathbf{Z}_2$ orientifold}
\label{secOZ2}

In this section we supplement our orbifold background by an O3
orientifold and show that in this case exotic instanton contributions do arise and
provide new terms in the superpotential. We refer to
{\it e.g.}~\cite{Douglas:1996sw,Berkooz:1996dw,Zwart:1997aj}
for a comprehensive discussion
of $\Ncal = 1$ and $\Ncal = 2$ orientifolds.

The first
ingredient we need is the action of the O3-plane on the various fields. Denote
by $\Omega$ the generator of the orientifold. The action of
$\Omega$ on the vertex operators for the various fields (ignoring for
the time being the Chan-Paton factors) is well known. The vertex
operators for the bosonic fields on the D3-brane contain, in the
0~picture, the following terms: $A_\mu \sim\ \partial_\tau x^\mu$ and
$\Phi^i \sim \partial_\sigma \bar z^i$. They both change sign under $\Omega$,
the first because of the derivative
$\partial_\tau$ and the second because the orientifold action for the
O3-plane is always accompanied by a simultaneous reflection of all the
transverse coordinates $z^i$.

The action of the orientifold on the Chan-Paton factors is realized by means of
a matrix $\gamma(\Omega)$ which in presence of an orbifold must satisfy the
following consistency condition
\cite{Douglas:1996sw}
\beq \gamma(g) \gamma(\Omega) \gamma(g)^T = +\, \gamma(\Omega)
\label{consistency}
\eeq
for all orbifold generators $g$.
This amounts to require that the orientifold projection commutes with the orbifold
projection. The matrix $\gamma(\Omega)$ can be either symmetric or
anti-symmetric. We choose to perform an anti-symmetric orientifold
projection on the D3 branes and denote the corresponding matrix by
$\gamma_-(\Omega)$. This requires having an even number $N_\ell$ of D3
branes on each node of the quiver so that we can write
\beq
 \gamma_-(\Omega) =  \begin{pmatrix}\epsilon_1 & 0 & 0   & 0   \cr  0 & \epsilon_2 & 0   & 0 \cr 0 & 0
& \epsilon_3   &  0 \cr  0   & 0   & 0 & \epsilon_4 \cr\end{pmatrix}
\label{easy}
\eeq
where the $\epsilon_\ell$'s are $N_\ell \times
N_\ell$ antisymmetric matrices obeying $\epsilon_\ell^2 = -1$. Using (\ref{chanpatonz2z2})
and (\ref{easy}) it is straightforward to verify that the consistency condition
(\ref{consistency}) is verified.

The field content of the stacks of fractional D3-branes in this orientifold model is
obtained by supplementing the orbifold conditions (\ref{orbaction}) with the
orientifold ones
\beq
A_\mu = - \gamma_-(\Omega) A_\mu^T
\gamma_-(\Omega)^{-1} ~~~,~~~  \Phi^l = - \gamma_-(\Omega) \Phi^{l T}
\gamma_-(\Omega)^{-1}.
\eeq
This implies that $A_\mu = \diag(A_\mu^1, A_\mu^2, A_\mu^3, A_\mu^4)$ with
$A_\mu^\ell = \epsilon_\ell A_\mu^{iT} \epsilon_\ell$.  Thus, the resulting gauge theory is a
$\mathrm{USp}(N_1) \times \mathrm{USp}(N_2) \times \mathrm{USp}(N_3) \times \mathrm{USp}(N_4)$
theory.
The chiral superfields, which after the orbifold have the structure (\ref{structure}),
are such that the $\Phi_{\ell m}$ component joining the nodes $\ell$ and $m$ of the quiver,
must obey the orientifold condition
$\Phi_{\ell m} = \epsilon_\ell \Phi_{m\ell}^T \epsilon_m$.
In the following, we will take $N_3=N_4=0$ so that
we are left with only two gauge groups and no tree level superpotential.

\subsection{Instanton sector}

Let us now consider the instanton sector, starting by analyzing the zero-mode
content in the neutral sector.
There are two basic changes to the previous story. The first is that the vertex operator for $a_\mu$
is now proportional to $\partial_\sigma x^\mu$, not to $\partial_\tau x^\mu$ and it remains
invariant under $\Omega$ (the vertex operator for $\chi_a$ still changes
sign). The second is that the crucial consistency condition discussed
in~\cite{Gimon:1996rq} requires that we now represent the action of
$\Omega$ on the Chan-Paton factors of the neutral modes by a symmetric
matrix which can be taken to be
\beq
\gamma_+(\Omega) =
\begin{pmatrix} 1 & 0 &0 & 0 \cr 0 & 1 & 0& 0 \cr 0&0&1&0 \cr 0 & 0 & 0 & 1 \cr
\end{pmatrix}~,
\label{gomega}
\eeq
where the $1$'s are $k_\ell \times k_\ell$ unit matrices. The matrix $a_\mu$
will be $4 \times 4$ block diagonal, {\it e.g.}
$a_\mu = \diag(a_\mu^1, a_\mu^2, a_\mu^3,a_\mu^4)$,  but now $a_\mu^\ell =
a_\mu^{\ell T}$. The most generic situation is to have a configuration with instanton numbers
$(k_1,k_2,k_3,k_4)$. By considering a configuration with $k_3 = 1$ and $k_1 =k_2 =k_4 =
0$, we can project out all bosonic zero-modes except for the four
components $a_\mu^3$ that we denote by $x_\mu$. The scalars $\chi^4\dots\chi^9$ are off-diagonal and we shall not
consider them further.

The nice surprise comes when considering the orientifold action on the
fermionic neutral zero-modes $M^{\alpha A}$ and $\lambda_{\dot \alpha
A}$. The orbifold part of the group acts on the spinor indices as in
(\ref{rg}), while the orientifold
projection acts as the reflection in the transverse space, namely
\beq
R(\Omega) = -i\,
\Gamma^{456789}
\label{romega}
\eeq
Putting together the orbifold projections (\ref{orbactionferm}) with the orientifold
ones
\beq
M^{\alpha A} = R^A_{~B}(\Omega)
\gamma_+(\Omega) (M^{\alpha B})^T \gamma_+(\Omega)^{-1}
~~~,~~~
\lambda_{\dot\alpha A} =
\gamma_+(\Omega) (\lambda_{\dot\alpha B})^T \gamma_+(\Omega)^{-1}R^B_{~A}(\Omega)
\label{orien}
\eeq
we can find the spectrum of surviving fermionic zero-modes. Using
(\ref{gomega}) and (\ref{romega}), it is easy to see that
(\ref{orien}) implies
\beq
M^{\alpha A} = (M^{\alpha A})^T ~~~,~~\lambda_{\dot\alpha A} = -
 (\lambda_{\dot\alpha A})^T~.
 \eeq
Thus, for the simple case where $k_3 = 1$ and $k_1 =k_2 =k_4 =
0$, all $\lambda$'s are projected out and only {\em two} chiral $M$ zero-modes remain:
$M^{\alpha---}$, to be identified with the
$\Ncal=1$ chiral superspace coordinates $\theta^\alpha$.

Also the charged zero-modes are easy to discuss in this simple
scenario. There are no bosonic modes
since the D-instanton and the D3-branes sit at different nodes while
the bosonic modes are necessarily diagonal. Most of the fermionic
zero-modes $\mu^A$ and $\bar\mu^A$ are also projected out by the orbifold
condition
\beq
\mu^A = R(g)^A_{~B} \gamma(g) \mu^B \gamma(g)^{-1}
 ~~~,~~~
 \bar\mu^A =
R(g)^A_{~B} \gamma(g) \bar\mu^B \gamma(g)^{-1}~.
\label{orbmu}
\eeq
Finally, the orientifold
condition relates this time the fields in the conjugate sectors,
allowing one to express $\bar\mu$ as a linear combination of the
$\mu$
\beq
\bar\mu^A = R(\Omega)^A_{~B} \gamma_+(\Omega) (\mu^B)^T
\gamma_-(\Omega)^{-1}~.
\label{orimuZ2}
\eeq
The only charged modes surviving these projections can be expressed, in block $4
\times 4$ notation, as
\beqs
\mu^2 = \begin{pmatrix} 0 & 0 & \mu_{13} & 0 \cr 0 &
0 & 0 & 0\cr  0& 0 & 0 & 0 \cr 0 & 0 & 0 & 0 \end{pmatrix},
 \quad \bar\mu^2 =
\begin{pmatrix} 0 & 0 & 0 & 0 \cr 0 & 0 & 0 & 0 \cr \bar\mu_{31} & 0 & 0 & 0 \cr 0 & 0 & 0 & 0
\end{pmatrix}, \nonumber \\ \mu^3 = \begin{pmatrix} 0 & 0 & 0 &0 \cr 0
& 0&\mu_{23} & 0\cr 0 & 0 & 0&0 \cr 0&0&0&0
\end{pmatrix}, \quad
\bar\mu^3 =
\begin{pmatrix} 0 & 0 & 0&0 \cr 0&0 & 0 & 0\cr 0&\bar\mu_{32} & 0 & 0 \cr 0&0&0&0
\end{pmatrix}~,
\label{muuuZ2}
\eeqs
where the entries, to be thought
of as column/row vectors in the fundamental/anti-fundamental of $\mathrm{SU}(N_\ell)$
depending on their position, are such that
$\bar\mu_{31} = - \mu_{13}^T \epsilon_1$ and $\bar\mu_{32} = -
\mu_{23}^T \epsilon_2$.

\begin{figure}[ht]
\begin{center}
{\includegraphics{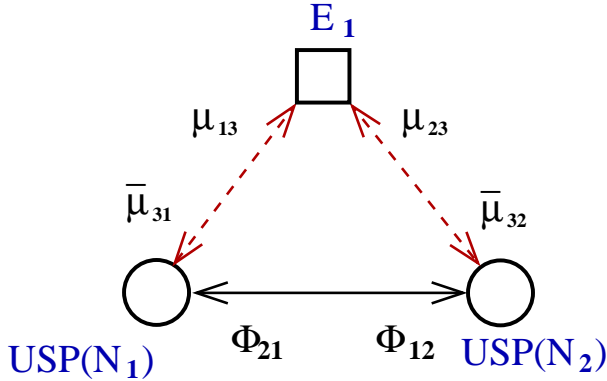}}
\caption{\small The generalized $\mathbf{Z}_2 \times \mathbf{Z}_2$ orientifold
quiver and the exotic instanton contribution.}
\label{quiverz2z2good}
\end{center}
\end{figure}

Thus, in the case where we have fractional D3 branes $(N_1, N_2, 0, 0)$ and an exotic
instanton $(0,0,1,0)$, the only surviving chiral field is $\Phi_{12}
\equiv \epsilon_1 \Phi_{21}^T \epsilon_2$, the orientifold projection
eliminates the offending $\lambda$'s and we are left with
just the neutral zero-modes $x_\mu$ and $\theta^\alpha$ and the
charged ones $\mu_{13}$ and $\mu_{23}$. This is summarized in the
generalized quiver of Fig.~\ref{quiverz2z2good}.

In this case the instanton partition function is
\beq
Z= \int
dx^4 d\theta^2 \,\,W
\eeq
where the superpotential $W$ is
\beq
W = \Ccal \int d\mu \,\,e^{-S_1 -S_2} =  \Ccal \int
d\mu_{13} d\mu_{23}\,\, e^{i \mu_{13}^T \epsilon_1 \Phi_{12} \mu_{23}}
~.
\eeq
This integral clearly vanishes unless $N_1=N_2$, in which case we
have
\beq
W~\propto~ \det (\Phi_{12})
\eeq
We thus see that exotic
instanton corrections are possible in this simple model.\footnote{
The gauge invariant quantity above can be rewritten as the Pfaffian
of a suitably defined mesonic matrix.}

It is interesting to note that the above correction is present in the
same case ($N_1=N_2\equiv N$) where the usual ADS superpotential for
$\mathrm{USp}(N)$ is generated~\cite{Intriligator:1995ne}
\beq
W_{ADS} =
\frac{\Lambda^{2N+3}}{\det (\Phi_{12})} \eeq
and its presence
stabilizes the runaway behavior and gives a theory with a non-trivial
moduli space of supersymmetric vacua given by $\det (\Phi_{12}) = \mathrm{const.}$
Of course, the ADS superpotential for this case can also be constructed along the
same lines as section~\ref{recoverADS}, see e.g.~\cite{Dorey:2002ik}. In fact,
this derivation is somewhat simpler than the one for the $SU(N)$ gauge group since
there are no ADHM constraints at all in the one instanton case.

We think the above situation is not specific
to the background we have been considering, but is in fact quite generic.
As soon as the $\lambda$ zero-modes are consistently lifted, we expect
the exotic instantons to contribute new superpotential terms.
As a further example, in the next section we will consider a
$\Ncal =2$ model, where exotic instantons will turn out to contribute to the prepotential.

\section{An $\Ncal = 2$ example: the  $\mathbf{Z}_3$ orientifold}
\label{sectz3}

\begin{figure}[ht]
\begin{center}
{\includegraphics{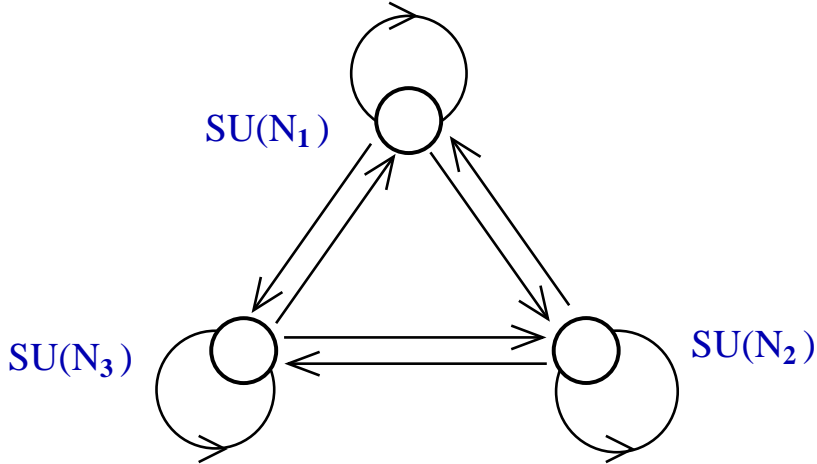}}
\caption{\small  The $\mathbf{Z}_3$ (un-orientifolded) theory. The lines with both ends on a single node represent adjoint
chiral multiplets which, together with the vector multiplets at each node constitute the $\Ncal = 2$
vector multiplets. Similarly, lines between nodes represent
chiral multiplets which pair up into hyper-multiplets, in $\Ncal =2$ language.}
\label{Z3}
\end{center}
\end{figure}

Let us now consider the quiver gauge theory obtained by placing an
orientifold O3-plane at a $\mathbf{C}\times \mathbf{C}^2/\mathbf{Z}_3$
orbifold singularity. In what follows we will use $\Ncal =1$ superspace notation.
We first briefly repeat the
steps that led to the constructions of such a quiver theory in the seminal
paper~\cite{Douglas:1996sw}.  Define $\xi = e^{2\pi i/3}$ and let the
generator of the orbifold group act on the first two complex
coordinates as
\beq
g: \begin{pmatrix} z^1 \cr z^2\cr\end{pmatrix} \to
\begin{pmatrix} \xi & 0  \cr 0 & \xi^{-1} \cr\end{pmatrix}
        \begin{pmatrix} z^1 \cr z^2\cr\end{pmatrix}~,
\eeq
while leaving the third one invariant. This preserves $\Ncal =
2$ SUSY. The action of the generator $g$ on the Chan-Paton factors
is given by the matrix
\beq
\gamma(g) =  \begin{pmatrix} 1 & 0 & 0
\cr 0 & \xi & 0\cr 0 & 0 & \xi^2 \cr \end{pmatrix}~.
\eeq
The $\Ncal =2$ theory obtained this way, summarized in Fig.~\ref{Z3},
is a three node quiver gauge theory with gauge groups $\mathrm{SU}(N_1)
\times \mathrm{SU}(N_2) \times \mathrm{SU}(N_3)$, supplemented by a cubic superpotential
which is nothing but the orbifold projection of the $\Ncal =4$ superpotential (its
precise form is not relevant for the present purposes).

As for the action of $\Omega$ on the Chan-Paton factors, we choose again to
perform the symplectic projection on the D3-branes. To do so, we must take $N_1$ to be even
and $N_2=N_3$, so that we can write
\beq
\gamma_-(\Omega) =  \begin{pmatrix} \epsilon & 0 & 0 \cr 0 & 0 & 1\cr
0 & -1 & 0 \cr \end{pmatrix}~,
\eeq
where $\epsilon$ is a $N_1 \times
N_1$ antisymmetric matrix obeying $\epsilon^2 = -1$ and the 1's
denote $N_2 \times N_2$ identity matrices. The matrices $\gamma(g)$
and $\gamma_-(\Omega)$ satisfy the usual consistency
condition~\cite{Gimon:1996rq,Douglas:1996sw} as in
(\ref{consistency}).

The field content on the fractional D3-branes at the singularity will be
given by implementing the conditions
\beqs
&&A_\mu = \gamma(g) A_\mu
\gamma(g)^{-1}~~~,~~~  \Phi^i = \xi^{-i} \gamma(g) \Phi^i
\gamma(g)^{-1}~,  \nonumber \\ &&A_\mu = - \gamma_-(\Omega) A_\mu^T
\gamma_-(\Omega)^{-1}~~~,~~~  \Phi^i = - \gamma_-(\Omega) \Phi^{i T}
\gamma_-(\Omega)^{-1}~.
\eeqs
The orbifold part of these conditions forces
$A_\mu$ and $\Phi^3$ to be $3 \times 3$ block diagonal
matrices, {\it e.g.} $A_\mu = \diag(A_\mu^1, A_\mu^2, A_\mu^3)$, while the
orientifold imposes that $A_\mu^1 = \epsilon A_\mu^{1T} \epsilon$ and
$A_\mu^2 = - A_\mu^{3 T}$. The resulting gauge theory is thus a
$\mathrm{USp}(N_1) \times \mathrm{SU}(N_2)$ theory. It is convenient, however, to still
denote $A_\mu^2$ and $A_\mu^3$ diagramatically as belonging to different
nodes with the understanding that these should be identified in the above
sense.

The projection on the chiral fields
can be done similarly and we obtain, denoting by $\Phi_{\ell m}$ the
non-zero entries of the fields $\Phi^1$ and $\Phi^2$ (only one can be
non-zero for each pair $\ell m$)
\beq
\Phi_{12} = -\epsilon \Phi_{31}^T,
\quad  \Phi_{13} = +\epsilon \Phi_{21}^T, \quad  \Phi_{23} =
\Phi_{23}^T, \quad  \Phi_{32} = \Phi_{32}^T~.
\eeq
The field content is
summarized in Table~2.
\begin{table} [ht]
\label{Z3fields}
\begin{center}
\begin{tabular}{c||c|c|}
& $\mathrm{USp}(N_1)$ & $\mathrm{SU}(N_2)$ \\
\hline
\hline  $\Phi_{12}$ & $\square$ & $\overline{\square} $ \\
\hline $\Phi_{21}$ & $\square$ & $\square$ \\
\hline $\Phi_{13}$ & $\square$ & $\square$ \\
\hline $\Phi_{31}$ & $\square$ & $\overline{\square}$ \\
\hline $\Phi_{23}$ & $\cdot$  & $\square \! \square$ \\
\hline $\Phi_{32}$ & $\cdot$ & $\overline{\square\! \square}$ \\
\hline
\end{tabular}
\caption{\small Chiral fields making up the quiver gauge theory.}
\end{center}
\end{table}

The theory we want to focus on in the following has rank assignment
$(N_1, N_2) = (0, N)$. This yields an $\Ncal=2$ $\mathrm{SU}(N)$ gauge theory
with an hyper-multiplet in the symmetric/(conjugate)symmetric
representation. We denote the $\Ncal=2$ vector multiplet by $\Acal$
whose field content in the block $3 \times 3$ notation is thus
\beq
\hat\Acal = \begin{pmatrix} 0 & 0 & 0
\cr 0 & \Acal & 0\cr 0 & 0 & -\Acal^T \cr \end{pmatrix}~.
\label{aca}
\eeq
In what follows we will be interested in studying corrections to the
prepotential $\Fcal$ coming from exotic instantons associated to
the first node (the one that is not populated by D3-branes). Let us then analyze
the structure of the stringy instanton sector of the present model, first.

\subsection{Instanton sector}

The most generic situation is to have a configuration with instanton numbers
$(k_1,k_2)$ (later we will be mainly concerned with a configuration with instanton
numbers $(1,0)$).

Let us start analyzing the zero-modes content in neutral sector.
The story is pretty similar to the one discussed in the previous section.
The vertex operator for $a_\mu$ is proportional to
$\partial_\sigma x^\mu$ and so it remains
invariant under $\Omega$. The action on the Chan-Paton factors of these
D-instantons must now be represented by a symmetric matrix which we take to
be
\beq
\gamma_+(\Omega) =
\begin{pmatrix} 1' & 0 & 0 \cr 0 & 0 & 1\cr 0 & 1 & 0 \cr
\end{pmatrix}
\eeq
where $1'$ is a $k_1 \times k_1$ unit matrix and
the $1$'s are $k_2 \times k_2$ unit matrices.

Because of the different orientifold projection, the matrices of
bosonic zero-modes behave slightly differently. The matrices $a_\mu$,
$\chi^8$ and $\chi^9$ will still be $3 \times 3$ block diagonal, {\it
e.g.}
$a_\mu = \diag(a_\mu^1, a_\mu^2, a_\mu^3)$,  but now $a_\mu^1 =
a_\mu^{1 T}$ and $a_\mu^2 = a_\mu^{3 T}$ whereas the same relations for
$\chi^8$ and $\chi^9$ will have an additional minus sign. The
remaining fields $\chi^{4\dots7}$ are off diagonal and we shall not
consider them further since we will consider only the case of one type
of instanton. By considering a configuration with $k_1 = 1$ and $k_2 =
0$, we can project out all bosonic zero-modes except for the four
components $a_\mu^1$ that we denote by $x_\mu$.

Let us now consider the orientifold action on the
fermionic neutral zero-modes $M^{\alpha A}$ and $\lambda_{\dot \alpha
A}$. The orbifold part of the group acts on the internal spinor
indices as a rotation
\beq
R(g) = e^{\frac{\pi}{3}\Gamma^{45}}
e^{-\frac{\pi}{3}\Gamma^{67}}~,
\eeq
while the orientifold acts through the matrix $R(\Omega)$ given in (\ref{romega}).
The orbifold and orientifold projections thus require
\beqs
&& M^{\alpha A} = R(g)^A_{~B} \gamma(g) M^{\alpha B} \gamma(g)^{-1}~~~,~~~
\lambda_{\dot\alpha A} =
\gamma(g) \lambda_{\dot\alpha B} \gamma(g)^{-1}R(g)^B_{~A} ~,
\\ && M^{\alpha A} = R(\Omega)^A_{~B}
\gamma_+(\Omega) (M^{\alpha B})^T \gamma_+(\Omega)^{-1}~~~,~~~
\lambda_{\dot\alpha A} =
\gamma_+(\Omega) (\lambda_{\dot\alpha B})^T \gamma_+(\Omega)^{-1}
R(\Omega)^B_{~A}~.\nonumber
\eeqs
Using the explicit expressions for the various matrices, we see that, for
the simple case where $k_1 = 1$ and $k_2 = 0$, all $\lambda$'s are
projected out and only {\em four} chiral $M$ zero-modes remain:
$M^{\alpha ---}$ and $M^{\alpha ++-}$ to be identified with the
$\Ncal=2$ chiral superspace coordinates $\theta^1_\alpha$ and
$\theta^2_\alpha$. Hence, also in this case the orientifold projection has cured the
problem encountered in section \ref{sectZ2Z2} (albeit in a $\Ncal = 2$
context now) and we can rest assured that the integration over the charged
modes will yield a contribution to the prepotential.

Let us now move to the charged zero-modes sector. Just as in the previous
model, there are no bosonic modes
since the D-instanton and the D3-branes sit at different nodes while
the bosonic modes are necessarily diagonal. Most of the fermionic
zero-modes $\mu^A$ and $\bar\mu^A$ are projected out by the orbifold
condition which is formally the same as in (\ref{orbmu}),
while the orientifold condition relates the fields in the conjugate sectors,
giving $\bar\mu$ as a linear combination of the
$\mu$'s according to
\beq
\bar\mu^A = R(\Omega)^A_{~B} \gamma_+(\Omega) (\mu^B)^T
\gamma_-(\Omega)^{-1}~.
\label{orimu}
\eeq
To summarize, the only
charged modes surviving the projection can be expressed, in block $3
\times 3$ notation as
\beqs
\mu^1 &=& \begin{pmatrix} 0 & 0 & 0 \cr 0 &
0 & 0\cr \mu & 0 & 0 \cr \end{pmatrix}~~~,~~~ \bar\mu^1 =
\begin{pmatrix} 0 & \mu^T & 0 \cr 0 & 0 & 0\cr 0 & 0 & 0 \cr
\end{pmatrix}~, \nonumber \\ \mu^2 &=& \begin{pmatrix} 0 & 0 & 0 \cr \mu'
& 0 & 0\cr 0 & 0 & 0 \cr \end{pmatrix}~~~,~~~ \bar\mu^2 =
\begin{pmatrix} 0 & 0 & -\mu'^T \cr 0 & 0 & 0\cr 0 & 0 & 0 \cr
\end{pmatrix}
\label{muuu}
\eeqs
where the entries are to be thought
of as column/row vectors in the fundamental/anti\-fun\-da\-men\-tal of $\mathrm{SU}(N)$
depending on their position.

\begin{figure}[ht]
\begin{center}
{\includegraphics{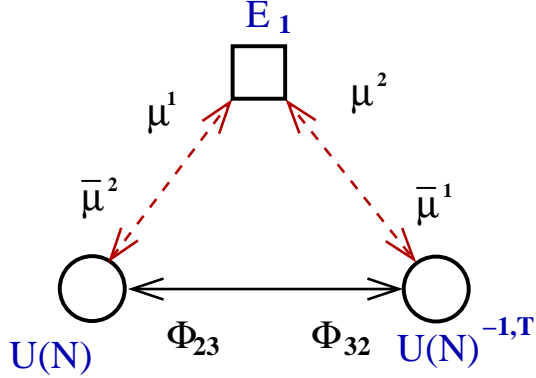}}
\caption{\small  The extended $\mathbf{Z}_3$ orientifold theory with $(0,N)$ fractional D3-branes
and $(1,0)$ instanton number. The
upper node (which would represent the $\mathrm{USp}(N_1)$ gauge group and disappears when we set
$N_1=0$ as in the case under consideration) is where the instanton sits. The lower nodes
denote only one gauge group. The charged
fermionic zero-modes follow Eq.~(\ref{muuu}). For simplicity we have not
drawn the lines denoting the adjoint.}
\label{OZ3}
\end{center}
\end{figure}

As anticipated, the configuration we want to consider is a $(0,N)$ fractional D3-branes
system together with an exotic $(1,0)$ instanton. The quiver structure, including the relevant
moduli, is depicted in Fig.~\ref{OZ3}.
It is now easy to see that inserting the expressions (\ref{aca}) and (\ref{muuu}) into
Eqs. (\ref{S1}), (\ref{S2})
and (\ref{master}) we finally obtain
\beq
Z =  \int dx^4 d\theta^4 \,\Fcal
\quad \mathrm{with} \quad \Fcal = \Ccal \int d\mu d\mu' \,\,e^{i \mu^T \Acal
\mu'}  \propto \det \Acal~.
\eeq
It would be interesting to study the potential implications of this result in the
gauge theory. There are many other simple models that could be analyzed along
these lines.

\section{Conclusions}

In this paper we have presented some simple examples of what seem to be
rather generic phenomena in the context of string instanton
physics. We paid particular attention to the study of the fermionic
zero-modes and their effects on the holomorphic quantities of the
theory. We have seen both examples where the instanton contributions vanish 
due to the presence of extra zero-modes and where they do not. In the second case, 
as explicitly shown in a $\Ncal = 1$ example, exotic instantons can have a 
stabilizing effect on the theory.

Although we have only considered some simple examples,
we would like to stress that these results are
quite generic and can be carried over to all orbifold gauge theories.
A future direction would be to try to be more systematic and analyze
the various possibilities encountered in more complex  $\Ncal=2$ and
$\Ncal = 1$ models. In a similar spirit, one should analyze the
multi-instanton contributions as well, since the total correction to
the holomorphic quantities will be the sum of all such terms. The
study of the zero-modes is expected to be even more relevant in this
case as it will probably make many contributions vanish.
With an eye to string phenomenology, one should also incorporate these
models into globally consistent compactifications and study the
effects of these terms there.

Lastly, it would be interesting to study the dynamical implications of
some of the terms generated. We briefly touched upon this at the
end of section \ref{secOZ2} when we mentioned the stabilizing effect of the exotic instanton
on the $\mathrm{USp}(N)$ theory. Although from the strict field theory point of
view these terms are thought of as ordinary polynomial terms in the
holomorphic quantities,\footnote{Save few (interesting) examples, these
terms are typically irrelevant and as a consequence should be naturally
suppressed by a high energy scale. Indeed, the terms generated by
stringy exotic instantons are suppressed by powers of the string scale.}
they are ``special'' when seen from the point
of view of string theory and they might therefore induce a particular
type of dynamics.

\section*{Acknowledgements}

We would like to thank many people for discussions and email exchanges
at various stages of this work that helped us sharpen the focus of the
presentation: M.~Bianchi, M.~Bill\`o, P.~Di~Vecchia, S.~Franco, M.~Frau,
F.~Fucito, S.~Kachru, R.~Marotta, L.~Martucci, F.~Morales, B.~E.~W.~Nilsson, D.~Persson,
I.~Pesando, D.~Robles-Llana, R.~Russo, A.~Tanzini, A.~Tomasiello, A.~Uranga, T.~Weigand and N.~Wyllard.

R.A., M.B. and A.L. are partially supported by the European Commission FP6
Programme MRTN-CT-2004-005104, in which R.A is associated to V.U. Brussel, M.B. to
University of Padova and A.L. to University of Torino.
R.A. is a Research Associate of the Fonds National de la Recherche
Scientifique (Belgium). The research of R.A. is also supported by IISN - Belgium
(convention 4.4505.86) and by the ``Interuniversity Attraction Poles Programme --Belgian Science Policy''.
M.B. is also supported by Italian MIUR under contract PRIN-2005023102 and by a
MIUR fellowship within the program ``Rientro dei Cervelli''.
The research of
G.F. is supported by the Swedish Research Council
(Vetenskapsr{\aa}det) contracts 622-2003-1124 and 621-2002-3884.
A.L. thanks the Galileo Galilei Institute for the hospitality and
support during the completion of this work.


\begin{thebibliography}{99}

\bibitem{Witten:1995gx} E.~Witten, 
Nucl.\ Phys.\  B {\bf 460} (1996) 541
  [arXiv:hep-th/9511030].  

\bibitem{Douglas:1995bn} M.~R.~Douglas, 
  arXiv:hep-th/9512077.  

\bibitem{Witten:1996bn} E.~Witten, 
Nucl.\ Phys.\  B {\bf 474}, 343 (1996)
  [arXiv:hep-th/9604030].  

\bibitem{Ganor:1996pe} O.~J.~Ganor,
Nucl.\ Phys.\  B {\bf 499}, 55 (1997)
  [arXiv:hep-th/9612077].  

\bibitem{Green:2000ke} M.~B.~Green and M.~Gutperle,
JHEP {\bf 0002} (2000) 014 [arXiv:hep-th/0002011].  

\bibitem{Billo:2002hm} M.~Billo, M.~Frau, I.~Pesando, F.~Fucito,
  A.~Lerda and A.~Liccardo, 
JHEP {\bf 0302} (2003) 045 [arXiv:hep-th/0211250].

\bibitem{Blumenhagen:2006xt} R.~Blumenhagen, M.~Cvetic and T.~Weigand,
  arXiv:hep-th/0609191.

\bibitem{Ibanez:2006da} L.~E.~Ibanez and A.~M.~Uranga, 
  arXiv:hep-th/0609213.  

\bibitem{Florea:2006si} B.~Florea, S.~Kachru, J.~McGreevy and
  N.~Saulina, 
  arXiv:hep-th/0610003.  

\bibitem{Abel:2006yk}
  S.~A.~Abel and M.~D.~Goodsell,
  arXiv:hep-th/0612110.

\bibitem{Akerblom:2006hx} N.~Akerblom, R.~Blumenhagen, D.~Lust,
  E.~Plauschinn and M.~Schmidt-Sommerfeld,
arXiv:hep-th/0612132.

\bibitem{Bianchi:2007fx} M.~Bianchi and E.~Kiritsis,
  arXiv:hep-th/0702015.  

\bibitem{Cvetic:2007ku} M.~Cvetic, R.~Richter and T.~Weigand,
  arXiv:hep-th/0703028.  

\bibitem{Bianchi:fu} M.~Bianchi, F.~Fucito, J.F.~Morales, arXiv:0704.0784 [hep-th].

\bibitem{Intriligator:2005aw} K.~Intriligator and N.~Seiberg, 
JHEP {\bf 0602} (2006) 031 [arXiv:hep-th/0512347].

\bibitem{Argurio:2006ew} R.~Argurio, M.~Bertolini, C.~Closset and
  S.~Cremonesi, 
JHEP {\bf 0609}, 030 (2006)
  [arXiv:hep-th/0606175].  

\bibitem{Argurio:2006ny} R.~Argurio, M.~Bertolini, S.~Franco and
  S.~Kachru, arXiv:hep-th/0703236.   

\bibitem{Dorey:2002ik} N.~Dorey, T.~J.~Hollowood, V.~V.~Khoze and
  M.~P.~Mattis, 
Phys.\ Rept.\ {\bf 371} (2002) 231 [arXiv:hep-th/0206063].

\bibitem{Bianchi:2007ft} M.~Bianchi, S.~Kovacs and G.~Rossi,
arXiv:hep-th/0703142.

\bibitem{Billo:2005fg}
M.~Billo, M.~Frau, S.~Sciuto, G.~Vallone, and A.~Lerda,
JHEP {\bf 0603} (2006) 023 [arXiv:hep-th/0511036].

\bibitem{Billo:2004zq}
M.~Billo, M.~Frau, I.~Pesando and A.~Lerda,
JHEP {\bf 0405} (2004) 023
[arXiv:hep-th/0402160];
M.~Billo, M.~Frau, F.~Lonegro and A.~Lerda,
JHEP {\bf 0505} (2005) 047
[arXiv:hep-th/0502084];
M.~Billo, M.~Frau, F.~Fucito and A.~Lerda,
 JHEP {\bf 0611} (2006) 012
 [arXiv:hep-th/0606013].

\bibitem{Witten:1995im} E.~Witten,
Nucl.\ Phys.\  B {\bf 460} (1996) 335
  [arXiv:hep-th/9510135].  

\bibitem{Atiyah:1978ri} M.~F.~Atiyah, N.~J.~Hitchin, V.~G.~Drinfeld
  and Yu.~I.~Manin, 
Phys.\ Lett.\  A  {\bf 65}, 185 (1978).  

\bibitem{Bertolini:2001gg} M.~Bertolini, P.~Di Vecchia, G.~Ferretti
  and R.~Marotta, 
  Nucl.\ Phys.\ B {\bf 630} (2002) 222 [arXiv:hep-th/0112187].

\bibitem{Bertolini:2003iv}
  M.~Bertolini,
  Int.\ J.\ Mod.\ Phys.\  A {\bf 18} (2003) 5647
  [arXiv:hep-th/0303160].

\bibitem{Franco:2005zu} S.~Franco, A.~Hanany, F.~Saad and
  A.~M.~Uranga, 
JHEP {\bf 0601} (2006) 011 [arXiv:hep-th/0505040].

\bibitem{Taylor:1982bp} T.~R.~Taylor, G.~Veneziano and
  S.~Yankielowicz, 
  Nucl.\ Phys.\  B {\bf 218} (1983)
  493.  

\bibitem{Affleck:1983mk} I.~Affleck, M.~Dine and N.~Seiberg,
  Phys.\  B {\bf 241} (1984) 493.  

\bibitem{Berenstein:2005xa}
  D.~Berenstein, C.~P.~Herzog, P.~Ouyang and S.~Pinansky,
  JHEP {\bf 0509} (2005) 084
  [arXiv:hep-th/0505029].

\bibitem{Bertolini:2005di}
  M.~Bertolini, F.~Bigazzi and A.~L.~Cotrone,
  Phys.\ Rev.\  D {\bf 72} (2005) 061902
  [arXiv:hep-th/0505055].


\bibitem{Friedan:1985ge} D.~Friedan, E.~J.~Martinec and S.~H.~Shenker,
Nucl.\  Phys.\  B {\bf 271} (1986) 93.  

\bibitem{Dixon:1986qv} L.~J.~Dixon, D.~Friedan, E.~J.~Martinec and
  S.~H.~Shenker, 
Nucl.\  Phys.\  B {\bf 282} (1987) 13.  

\bibitem{DiVecchia:1997pr}
  P.~Di Vecchia, M.~Frau, I.~Pesando, S.~Sciuto, A.~Lerda and R.~Russo,
  Nucl.\ Phys.\  B {\bf 507} (1997) 259
  [arXiv:hep-th/9707068];
  M.~Bertolini, P.~Di Vecchia, M.~Frau, A.~Lerda, R.~Marotta and I.~Pesando,
  JHEP {\bf 0102} (2001) 014
  [arXiv:hep-th/0011077];
M.~Bertolini, P.~Di Vecchia, M.~Frau, A.~Lerda and R.~Marotta,
  Nucl.\ Phys.\  B {\bf 621} (2002) 157
  [arXiv:hep-th/0107057].

\bibitem{DiVecchia:1999rh} P.~Di Vecchia and A.~Liccardo,
  NATO Adv.\ Study Inst.\ Ser.\ C.\ Math.\
  Phys.\ Sci.\  {\bf 556} (2000) 1 [arXiv:hep-th/9912161].

\bibitem{Martucci:2005rb}
  L.~Martucci, J.~Rosseel, D.~Van den Bleeken and A.~Van Proeyen,
  Class.\ Quant.\ Grav.\  {\bf 22}, 2745 (2005)
  [arXiv:hep-th/0504041].

\bibitem{Bergshoeff:2005yp}
  E.~Bergshoeff, R.~Kallosh, A.~K.~Kashani-Poor, D.~Sorokin and A.~Tomasiello,
  JHEP {\bf 0510}, 102 (2005)
  [arXiv:hep-th/0507069].
\bibitem{Pradisi:1988xd} G.~Pradisi and A.~Sagnotti,
Phys.\ Lett.\  B {\bf 216} (1989) 59.

\bibitem{Gimon:1996rq} E.~G.~Gimon and J.~Polchinski,
Phys.\ Rev.\  D {\bf   54} (1996) 1667 [arXiv:hep-th/9601038].

\bibitem{Douglas:1996sw} M.~R.~Douglas and G.~W.~Moore,
arXiv:hep-th/9603167.

\bibitem{Berkooz:1996dw} M.~Berkooz and R.~G.~Leigh,
 Nucl.\ Phys.\  B {\bf 483} (1997) 187
 [arXiv:hep-th/9605049].  

\bibitem{Zwart:1997aj} G.~Zwart,
Nucl.\ Phys.\  B {\bf 526} (1998) 378
  [arXiv:hep-th/9708040].  

\bibitem{Intriligator:1995ne} K.~A.~Intriligator and P.~Pouliot,
Phys.\ Lett.\  B {\bf
  353} (1995) 471 [arXiv:hep-th/9505006].  

\end{thebibliography}
\end{document}